# When Energy-Efficiency May Not Yield Positive Climate Impact – The Case of Adaptive Radiative Coolers

Nithin Jo Varghese,[1,3†] Jyothis Anand,[2†] Jyotirmoy Mandal[1,3,*]

[1] Department of Civil & Environmental Engineering, Princeton University, Princeton, USA
[2] Buildings and Transportation Science Division, Oak Ridge National Laboratory, Oak Ridge, USA
[3] Princeton Materials Institute, Princeton University, Princeton, USA
[†]These authors contributed equally to the work
*Corresponding author: Jyotirmoy Mandal (jm3136@princeton.edu)

## Abstract

The climate impact of building envelopes is often quantified using their energy savings and $CO_2$ emission reduction benefits. However, building envelopes also trap solar and thermal infrared heat, which is dissipated as a direct heating penalty into our warming planet. For static or adaptive envelopes that passively heat buildings by radiatively retaining heat, these two effects are antagonistic. Yet, their net effect remains unexplored.

In this study, we compare the emission reductions benefit, and direct heating penalty of two classes of roof envelopes, traditional and adaptive radiative coolers (TRCs and ARCs). Calculations for buildings in different urban climates show that relative to TRCs like cool roofs, ARCs like smart roofs may have a net heating impact on earth well past this century. Thus, despite their relative energy savings and $CO_2$ emissions reductions benefits, adaptive envelopes on roofs have a negative climate impact relative to traditional cooling designs. Our findings are generalizable across climates and a range of building envelopes and call for a rethinking of how sustainability is quantified for building envelopes, and of material and architectural design for buildings.

## Introduction

Radiative cooling of sky-facing terrestrial surfaces involves heat loss to the sky at wavelengths where the atmosphere is transparent (and reflection of solar wavelengths during the daytime). The zero-energy, zero-carbon functionality of this spontaneous process has seen it increasingly explored for passively cooling environments at local[1,2] and global[3–5] scales. Cool roof coatings are perhaps the most widely known and longest-used example of radiative coolers,[6,7] and recent years have seen the development,[8–13] field demonstrations,[14–16] and theoretical explorations of the cooling capabilities of solar reflective, thermally emissive radiative coolers – much of it geared towards their biggest application, cooling roofs of buildings.[7,12,15] However, a significant concern is the potential overcooling of buildings by radiative coolers in cold weather – traditional radiative coolers (TRCs) are by design static, and their persistent heat loss to space may raise heating demand, and associated costs and $CO_2$ emissions, when the weather is cold (**Figure 1**).[1]

To address this issue, researchers have developed thermochromic,[17,18] electrochromic,[19,20] fluidic,[21] and other[22] designs that can be passively or actively tuned to thermoregulate buildings across different weather conditions. Like traditional radiative coolers, these adaptive radiative coolers (ARCs) have a high solar reflectance ($R_{solar}$, across $\lambda \sim 0.3 - 2.5\ \mu m$), and a high thermal infrared (TIR, $\lambda \sim 2.5 - 40\ \mu m$) emittance ($\varepsilon_{IR}$) that subtends the longwave infrared (LWIR, $\lambda \sim 8 - 13\ \mu m$) atmospheric transmission window during warm weather. This enables radiative cooling to the sky. During cold weather, however, ARCs switch to heating mode by lowering their $R_{solar}$ and/or $\varepsilon_{IR}$. This enables them to capture solar heat, and/or minimize skywards LWIR heat loss, and passively keep buildings warmer than traditional radiative coolers would. This can lower building heating loads, and corresponding $CO_2$ emissions and costs – a significant advantage over TRCs. Naturally, ARCs have witnessed a high research interest from a materials perspective, and have been proposed as a frontier in radiative cooling research.[23]

A driving motivation for radiative cooling building envelopes, whether traditional or adaptive, is their promise for sustainable thermoregulation of buildings in the context of climate change. To date, this has been widely explored using physical models which quantify and show that ARCs reduce annual energy usage and $CO_2$ emissions in buildings relative to TRCs, making them an apparently more sustainable option [17,24]. Such analyses, however, overlook the direct radiative impact of TRCs and ARCs on the environment – in other words, how much heat is directly sent by a TRC or ARC from earth to space, in addition to the indirect cooling impact on earth through $CO_2$ emissions reductions. A consideration of this immediately raises a complex issue: in cold weather, ARCs have two opposing thermal impacts on the environment. As stated above, they reduce heating-related $CO_2$ emissions, which is desirable from a climate change perspective. However, to do so, they trap solar and/or TIR heat, thus replacing a part of the earth's surface that was covered with a radiative cooler sending heat to space, with a surface that traps heat (**Figure 1**). Crucially, the trapped low-grade heat is dissipated into the environment, warming the earth.

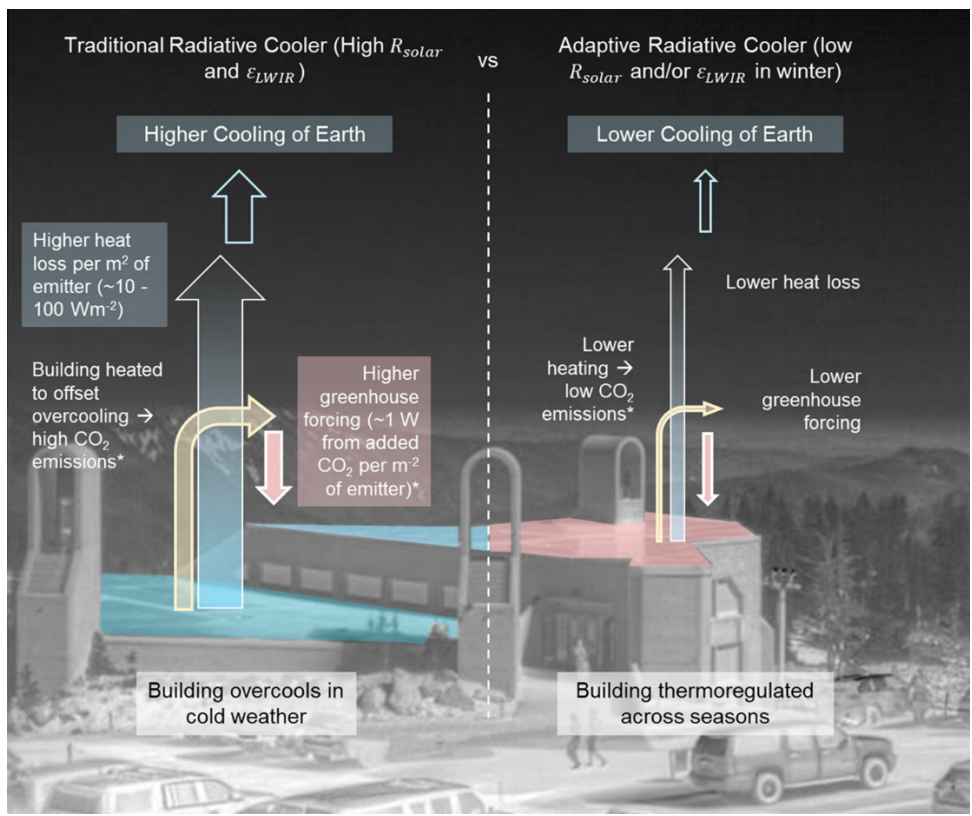

**Figure 1.** Schematic showing the building-level and direct thermal impact of (left) traditional and (right) $\varepsilon_{IR}$-switching adaptive radiative coolers. The former radiates heat originating from buildings and the environment to space across seasons, cooling the earth, but overcools buildings during cold weather and increases $CO_2$ emissions related to building heating loads. Adaptive radiative coolers passively keep building warmer in cold weather and reduce $CO_2$ emissions, but only by reducing heat loss to space from the part of the earth it covers. The thermal image was taken on Mount Hood, USA, 45.331°N, 121.711°W.

\* Indicates that for heating buildings in cold weather, fossil fuel rather than renewable energy-based heating was assumed.

Despite its profound implications, the relationship between and consequence of these two antagonistic impacts remain unexplored. The impact of optically functional envelopes on buildings and the environment is an active topic in the building and urban sciences[23,25,26] with implications for both industry and policy[27–29]. Concurrently, there has been a parallel and growing scrutiny of carbon footprints of building technologies, materials and construction practices[30,31]. Against this backdrop, there has been tremendous efforts towards material design [32] and commercialization[33,34] of TRCs and ARCs, with sustainable thermoregulation as a goal. Given that ARCs have captured the scientific[23,35–40] and public imagination[41,42], it is thus timely to consider their holistic impact on the environment.

In this paper, we explore whether adaptive radiative coolers have a more desirable climate impact than traditional radiative coolers, by modelling their relative $CO_2$ emissions reductions, or the operational cooling benefit, and their radiative heating impact on earth, or the operational thermal penalty. Using EnergyPlus™ and MODTRAN®, and assumptions that favor emissions reductions by ARCs, we calculate these quantities for 15 US cities – Albuquerque, Atlanta, Baltimore, Boulder, Chicago, Duluth, Fairbanks, Helena, Houston, Las Vegas, Miami, Minneapolis, Phoenix, San Francisco, and Seattle – representing different climate zones, and for different climate scenarios. We find that across climates, ARCs have a net heating penalty on the earth in the near term and yield a net cooling benefit only after a century or more in all but perpetually cold locations. Other roof envelopes, like dark or metallic roofs, have even greater penalties.

Additionally, we provide physical explanations of the varying penalties of ARCs in different cities, which could be generalized to locations beyond the cities we studied. Lastly, we explore how roofs can be designed to simultaneously employ TRCs and ARCs to benefit both ambient and indoor environments. Our results call for a rethinking of how sustainability is quantified for roof envelopes in general, and of material and architectural design for buildings. More broadly, it adds to a body of literature that have more holistically evaluated the climate impact of energy technologies or surface interventions [43–45].

## Calculation of the Net Operational Thermal Footprints of TRCs and ARCs

To calculate the net operational thermal footprint of TRCs and ARCs on roofs of individual buildings, we accounted for both direct radiative heat transfer from roofs to space, and indirect impact through change in $CO_2$ emissions *at the single-building level* (**Figure 2A**). The term 'operational' is used to distinguish our calculations from non-operational aspects of the life cycle of TRCs and ARCs, like manufacturing and installation. Unless otherwise stated, all associated variables are assumed to be operational as well. We performed our calculations for mid-rise buildings with flat roofs in 15 US cities representing different climate zones. Building-level data, such as construction and architecture, was obtained from DOE standard models, and the hourly weather data was obtained from Typical Meteorological Year projections[46,47]. Further details about the data are provided in the SI, Section 1.

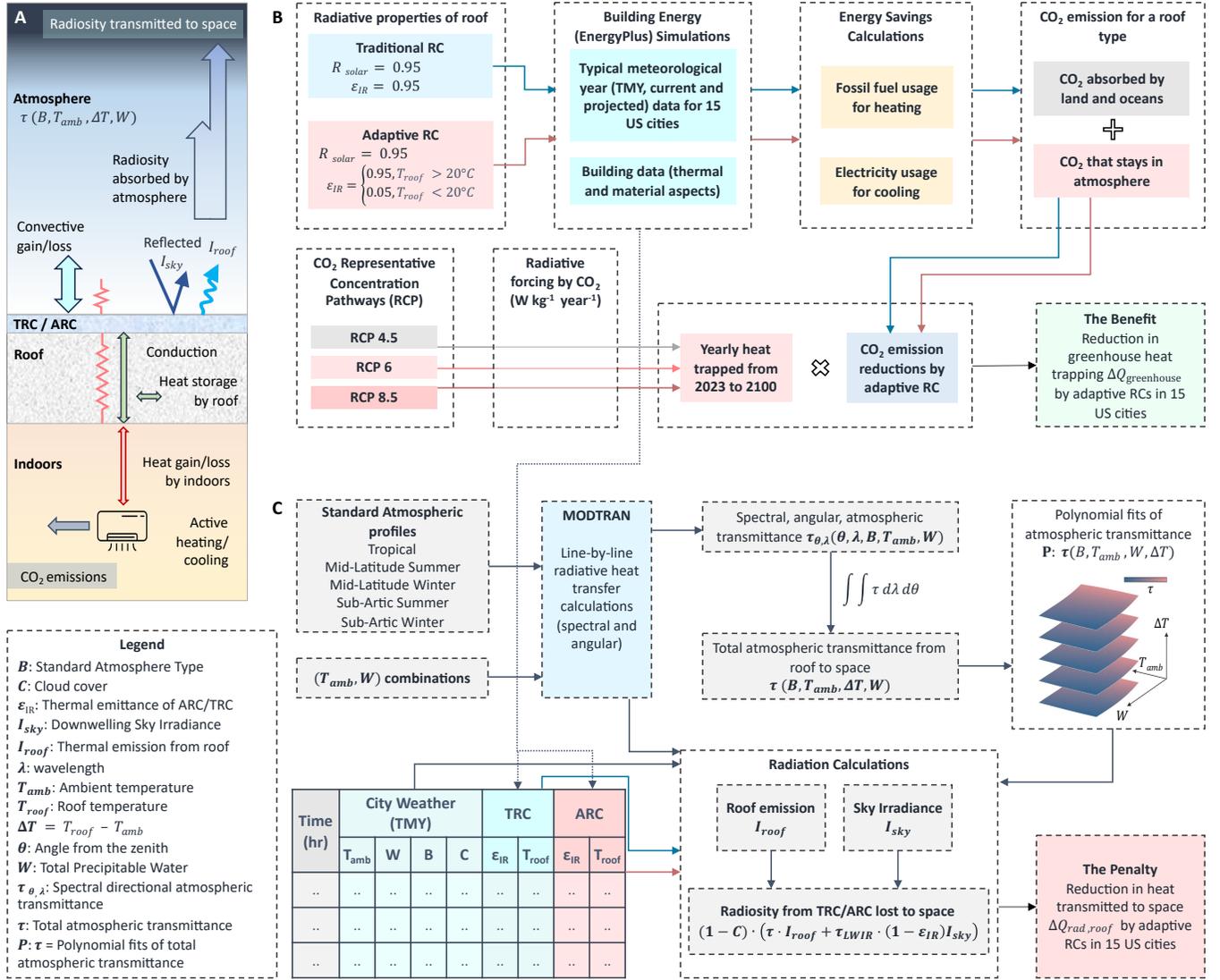

**Figure 2.** (**A**) A schematic showing the physical mechanisms considered in our analysis. (**B**) Flowchart detailing the calculation of energy savings and $CO_2$ emissions reductions by TRCs and ARCs. (**C**) Flowchart detailing the procedure to calculate the heat transmitted to space from TRC and ARC roofs.

For our calculations, we assumed that roofs had an unobstructed view of the sky and were covered entirely by a TRC or an ARC. The TRC was assumed to have an isotropic solar reflectance $R_{solar}$=0.95 and thermal emittance $\varepsilon_{IR}$=0.95. For calculations involving ARCs, we focused on emittance switching designs and modelled solar adaptive ARCs that switch $R_{solar}$ and dual-mode ARCs that switch both $\varepsilon_{IR}$ and $R_{solar}$ for a subset of the cities (SI, Section 5). Unless stated otherwise, we assumed that emittance switching ARCs have high $R_{solar}$=0.95, and $\varepsilon_{IR}$=0.95 when they are warmer than 20°C and $\varepsilon_{IR}$=0.05 when cooler. Any sunlight not reflected was assumed to be absorbed by the TRC or ARC. In other words, the ARCs generally had the same radiative cooling functionality as TRCs during warm weather, and a relative heating functionality in cold weather. For reasons that we elaborate later, we did not model static envelopes that are dark or non-emissive. **Figure 2** shows a graphical summary and flowcharts of our calculations, which are detailed below.

## The Benefit: $CO_2$ Emissions Reduction by Adaptive Radiative Coolers in Cold Weather

The magnitude of cold-weather heating of buildings by ARCs relative to TRCs on roofs determines the reduction in heating loads, and since we assume buildings are heated by burning natural gas, a corresponding reduction in $CO_2$ emissions. There is also a smaller, thermal inertia-induced effect when the roof is close to the optical transition temperature, when the ARC cools and switches to a low-ε mode even as underlying depths of the roof remains warm. This increases the electrical cooling load, which we also assume to come from fossil fuels[48]. Since $CO_2$ is a greenhouse gas, the net emissions reductions lead to a reduction in heat trapping by the atmosphere. We term this reduced heat trapping as the operational "benefit" ARCs have over TRCs, and denote it as $\Delta Q_{greenhouse}$. Notably, this assumes that the reduction in heat trapping is not local (even when it briefly is at the moment and location of $CO_2$ reduction, it is too small relative to the effect of $CO_2$ in the local atmosphere), and is distributed into the $CO_2$ reservoir of the atmosphere.

We calculated the building heating and cooling energy consumption for TRC and ARC roofs by entering data on building thermal properties, energy supply parameters, location, hourly TMY parameters, and the optical properties of the TRC or ARC into EnergyPlus™ (**Figure 2B**). The fossil fuel and electricity consumptions yielded by the finite difference thermal calculations in EnergyPlus™ was used to calculate $CO_2$ emissions for each year from 2023 to 2100 (SI, Section 1). Relevant energy to $CO_2$ conversion factors of ~0.053 kg/$MJ$ for natural gas usage for heating [49,50], and 0.4 kg/kWh for electricity usage for cooling [51], was used. The total $CO_2$ emissions from a building was the sum of the contributions from natural gas and electricity usage. The difference in kilograms of $CO_2$ emissions per $m^2$ roof area between the ARC and TRC cases was the emissions reduction by the ARC roof. The effective emissions reduction, however, was estimated to be 45% of that value, as the rest gets absorbed from the air by the oceans and land [52,53].

The impact of the effective emissions reductions ($dC$) on radiative forcing $F$ (W·m$^{-2}$) at a given time $t$ depends on the concentration $C$ (ppm) of atmospheric $CO_2$, where 1 ppm corresponds to a mass of 7.829 x 10[12] kg of $CO_2$ distributed throughout the atmosphere. This can be derived from a well-known relationship[54]:

$$\int dF(t) = \int \frac{5.35}{C(t)} dC \qquad\qquad 1$$

Therefore, to calculate the impact of effective $CO_2$ emissions reductions over a given number of years from 2023, we considered different representative $CO_2$ concentration pathways, RCP 8.5, 6 and 4.5. RCPs are projections of the future trends in greenhouse gas emissions and concentration until the year 2100[55]. RCP 4.5, 6, and 8.5 represent increasing levels of $CO_2$ emissions (and hence atmospheric concentrations) leading to global warming, with RCP 8.5 in particular representing projections based on current and stated policies.[53] For each scenario, we calculated $dF/dC$ for every year from 2023 to 2100 per kg reduction in $CO_2$ emissions, and multiplied it by the effective emissions reductions by ARCs. Importantly, since atmospheric $CO_2$ has a lifetime of ~300 years, emissions reductions in any year between 2023 and 2100 would impact forcing in all subsequent years of that period. We accounted for this cumulative effect. For our primary analysis, we focused on RCP 8.5 (business as usual scenario), and considered two climate projections, one where we assumed the current TMY throughout the period of the simulation, and one where we use a projection by Chowdhury et. al[46]. Results based on calculations using the latter is presented in **Figure 3.** Details of the calculations for RCPs 6 and 4.5 are presented in the SI, Section 5. Finally, the benefit $\Delta Q_{greenhouse}$, the reduction in greenhouse heat trapping by ARCs relative to TRCs, was calculated by multiplying yearly values of $dF$ with time, in W yr m$^{-2}$, for the 15 US cities we chose (**Figure 2B**). Similar calculations were performed for solar-adaptive and dual mode ARCs (Supporting Information, Section 5).

## The Penalty: Reduction in Radiative Heat Loss to Space from Adaptive Radiative Coolers

An ARC, owing to its low $\varepsilon_{IR}$ when cold, radiates less heat $I_{roof} = \varepsilon_{IR}\sigma T_{roof}^4$ than a TRC, but partially compensates for this by reflecting more of the downwelling sky radiation $I_{sky}$ upwards (**Figure 2A**). It should be noted that we assume the roof is optically opaque. A significant fraction of the upwelling radiosity $I_{roof} + (1 - \varepsilon_{IR})I_{sky}$ from a TRC or ARC roof is transmitted to space, and the rest absorbed, and partially reemitted to space, by the atmosphere.[57] The reduction in this radiosity as seen from space is the heating impact on earth. A conservative estimate of this reduction – which we call the operational penalty – is the component that would be directly transmitted through the atmosphere to space, and is modulated by the atmospheric transmittance $\tau$ [57]. Importantly, any convective and radiative heating of the ambient air arising from the reduced radiosity from a roof dissipates as low-grade heat into the thermal reservoir of the atmosphere and is trapped on earth.

**Figure 2C** shows how the penalty was calculated. While the calculation of $I_{roof}$ of TRCs and ARCs, using $\varepsilon$ and roof temperature $T_{roof}$ derived from EnergyPlus™ (**Figure 2B**) was fairly straightforward, calculation of spectrally and meteorologically sensitive $\tau$ was less so. We used MODTRAN® to calculate hemispherical $\tau(B, T_{amb}, W, \lambda)$ for different atmosphere types ($B$, namely tropical, sub-tropical summer and winter, sub-Arctic summer and winter), ambient temperatures ($T_{amb}$), and total precipitable water ($W$) levels over the thermal wavelengths ($\lambda \sim 0.3 - 40 \ \mu m$). Since our TRCs and ARCs were spectrally flat emitters, this was used to derive the total atmospheric transmittance $\tau(B, T_{amb}, W, T_{roof})$ for different roof temperatures ($T_{roof}$). Thermal emission from the roof to space $\tau(B, T_{amb}, W, T_{roof}) \cdot I_{roof}(\varepsilon_{IR}, T_{roof})$ was then calculated from TMY and EnergyPlus™ data. Where cloud cover $C$ blocked direct heat transmission, $\tau$ was weighed by $1 - C$.

$I_{sky}$ was calculated from radiative sky temperatures in the TMY database and accounted for irradiances from both clear-skies and clouds. However, since the overwhelming fraction of roof radiosity transmitted to space occurs in the LWIR atmospheric window, it was really $I_{sky,LWIR}$, the LWIR part of $I_{sky}$, that was important. Using MODTRAN® we calculated $I_{sky,LWIR}$. The fraction of it reflected by the TRC or ARC roof, and then transmitted to space, was then calculated as $(1 - C) \cdot \tau_{LWIR}(B, T_{amb}, W) \cdot (1 - \varepsilon_{IR})I_{sky,LWIR}$. Notably, $\tau_{LWIR}$ here was independent of $T_{roof}$. The total radiosity, then, was:

$$I_{rad,roof} = (1 - C)[\tau(B, T_{amb}, W, T_{roof}) \cdot I_{roof}(\varepsilon_{IR}, T_{roof}) + \tau_{LWIR}(B, T_{amb}, W) \qquad 2$$
$$\cdot (1 - \varepsilon_{IR})I_{sky,LWIR}]$$

The penalty for the $i$th year from 2023-2100 was calculated based on the projected TMY data for that year, and then used to calculate the penalty $\Delta Q_{rad,roof}$ up to the $Y$th year in that period, as follows:

$$\Delta Q_{rad,roof,i} = (I_{rad,TRC\ roof} - I_{rad,ARC\ roof}) \cdot 1 \implies \Delta Q_{rad,roof} = \sum_{i=1}^{Y} \Delta Q_{rad,roof,i} \qquad 3$$

For solar-adaptive and dual-mode ARCs (Supporting Information, Section 5), $\Delta Q_{rad,roof}$ was similarly calculated by accounting for the solar and solar-to-TIR radiosities respectively.

We note that since ARCs modulate $I_{roof}$ to thermoregulate building indoors, whether the penalty is in heat lost from the terrestrial environment, or in actively generated heat from indoors, may not be apparent. As shown in **Figure 2A**, the thermal resistance between the TRC/ARC and the outdoor environment is much lower than that between TRC/ARC roof surface and the insulated indoors, meaning that heat flows to/from indoors have little impact on $I_{roof}$. Indeed, heat fluxes from EnergyPlus™ show that compared to convective and radiative heat flows from the environment, heat flow from the indoors to the roof surface is ~10-20x smaller (Supporting Information, Section 9). The penalty $\Delta Q_{rad,roof}$ thus primarily impacts the environment.

## Results – the Net Operational Thermal Footprint of ARCs relative to TRCs

The above calculations enable us to compare the emissions reduction benefit of ARC roofs relative to TRC roofs for three RCPs with the radiative penalty, for 15 US cities. **Figure 3** shows the plots of the net penalties accumulated by an ARC roof relative to a TRC roof in each as a function of time. It is clear that in the foreseeable future, in all the cities we studied, ARCs will have a net heating impact on the environment. For RCP 8.5 scenario, which is closest to our assumption that fossil fuels will continue to be used for heating buildings in the future, ARCs have net penalties beyond 2100 for all the cities we modelled.

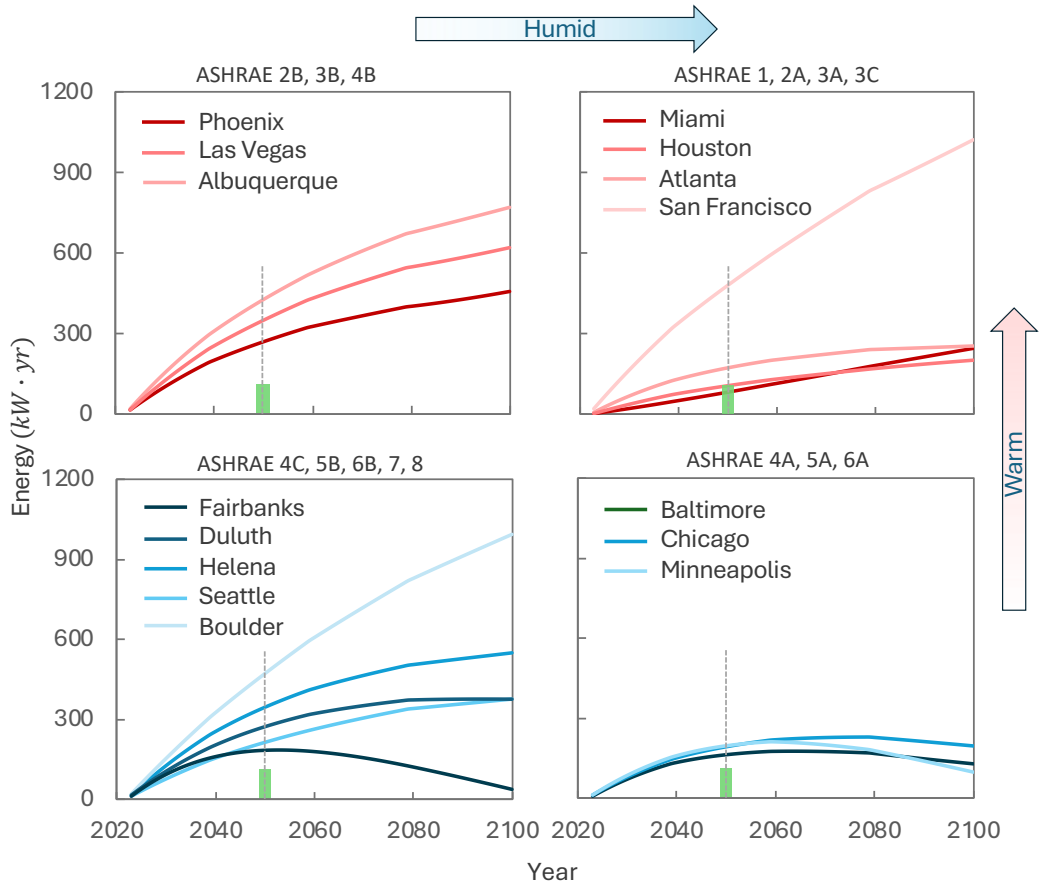

**Figure 3.** Time evolution of the net penalties by a mid-sized residential building (of roof area 784 m²) deploying an adaptive RC on the roof over a traditional RC in fifteen US cities grouped according to ASHRAE climate zones for buildings. The penalties were calculated using the TMY projections, except for Fairbanks, for which the current TMY data was used due to projections data not being available. Compare the net penalty clocked by the year 2050 (grey dashed line), with the thermal penalty (green bar) due to $CO_2$ emissions, of air conditioning an average US household (of area 169 m²).

The penalties that we observe are significant. For each city in **Figure 3,** the intersection of the grey dotted line with the curve represents the net penalty (($\Delta Q_{rad,roof} - \Delta Q_{greenhouse}$) × $Roof\ Area$) accrued by the year 2050, i.e., 28 years from the start of our calculations (2023), corresponding to the UN 2050 net zero target. Depending on the city, ARCs trap a large amount of heat on earth relative to TRCs, between ~105 to ~612 W·yr·m⁻², which is equivalent to heat trapped by ~0.88 to ~5.15 kg of $CO_2$ yr⁻¹ over the same period. For the large roofs we modelled (~784 m² area), this corresponds to ~82 to ~480 kW·yr, or an effective $CO_2$ emission of ~690 to ~4038 kg yr⁻¹. To put these values into perspective, $CO_2$ emissions from air conditioning consumption over the same period in an average household (of area ~ 169 m²) is 926 kg yr⁻¹, or the green bars corresponding to 110 kW·yr in **Figure 3**[58]. The net penalties accumulated by the ARCs relative to the TRCs are thus significant, and crucially, similar to those of the active thermoregulation methods ARCs and TRCs are supposed to replace.

In addition to studying the relative impact of ARCs in different cities with diverse climates in **Figure 3**, we also analyzed the sensitivity of our findings to different modelling parameters. The results are presented in the next section and indicate the robustness of our findings.

**Figure 3** categorizes the cities in our simulations by temperature and humidity according to ASHRAE climate zones for buildings[59]. As evident, for more humid regions, the penalties are generally lower. However, even within categories, there are significant variations in the magnitudes and long-term trends of the penalties. For example, Fairbanks and Boulder (Cold, dry) have different magnitudes, while Miami and Atlanta (warm, humid) have different trajectories. A geographical snapshot of the net penalties in 2050 (**Figure 4**) reveals a similarly complicated picture. These indicate that the magnitude and trends of the penalties are not explainable by location or existing climate zone classifications (e.g. ASHRAE) alone. This led us to consider the geographic and climatic idiosyncrasies of each city. For instance, Boulder, Albuquerque, Las Vegas, Phoenix, and San Francisco all have winters characterized by low humidity, cloudless skies, and moderately cold air temperatures $T_{amb}$ (SI, Section 2). The first two of these heightens $(1 - C) \cdot \tau$ and lowers $I_{sky}$, while the last convectively heightens $T_{roof}$ and thus $I_{roof}$. Thus, $\Delta Q_{rad,roof}$ between ARCs and TRCs are quite high. In contrast, in cities like Duluth, Minneapolis and Chicago, where the winters are very cold (Supporting Information, Section 3), a low $T_{roof}$ means that any change in $I_{rad,roof}$ is due to a change in $\varepsilon$ (and thus the penalty) is smaller. Thus, the net penalty is smaller as well.

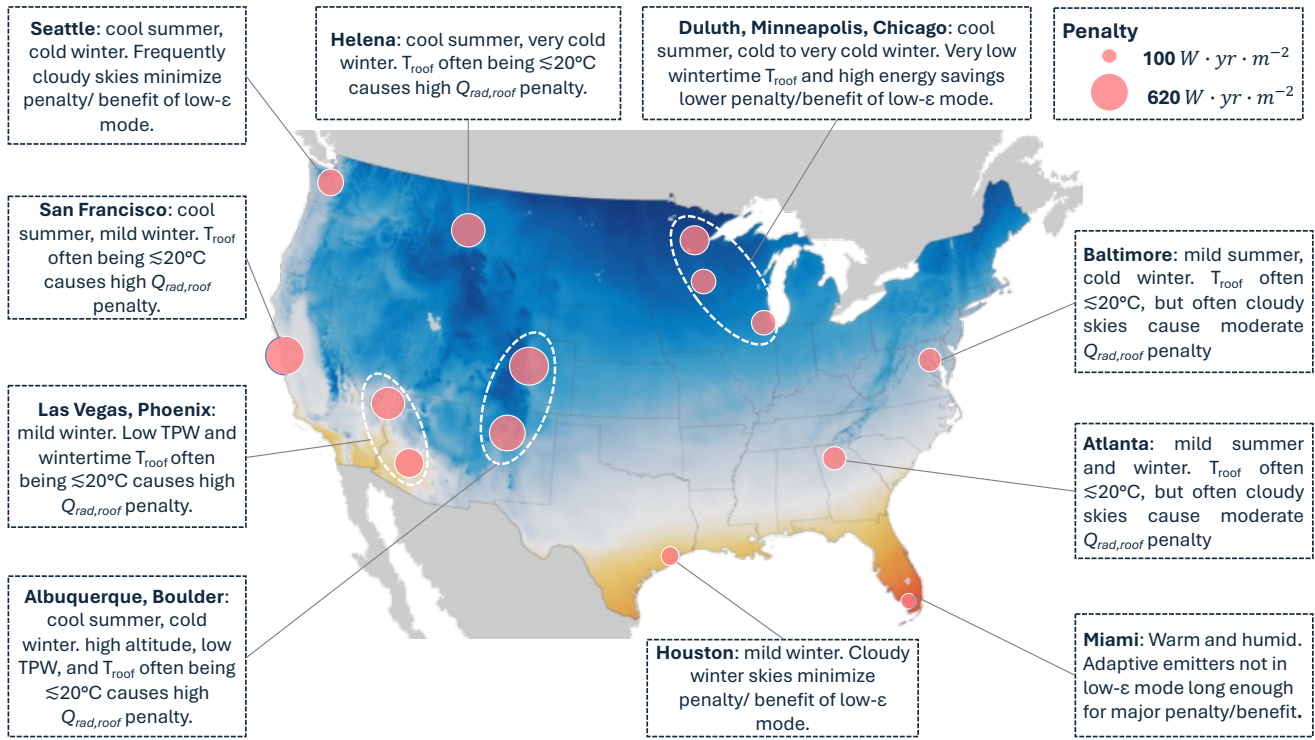

**Figure 4.** A geographical snapshot of the relative impact of ARCs in 2050 for simulated cities in the contiguous US, with the background color qualitatively representing average wintertime temperatures from 1991-2020. Physical explanations for the magnitude of the penalties are provided.

Intriguingly, for cities with milder winters but otherwise similar climate conditions as cities with harsher winters, the penalties are not very different (e.g. Minneapolis, Chicago, Atlanta). This is likely because while milder winters are shorter, they are also often warmer. Thus, even though ARCs stay in the heating mode for a shorter time in such cities, a higher $\Delta Q_{rad,roof}$ makes up for it. This indicates that even if winters grow shorter and milder with climate change, ARCs would continue to have significant penalties.

The most powerful determinant of $\Delta Q_{rad,roof}$ is cloud cover. Seattle, for instance, has a warmer winter than Duluth, Minneapolis and Chicago, but cloudy winter skies (Supporting Information, Fig. S17) that prevent heat loss to space and reduce $\Delta Q_{rad,roof}$. Houston, which has a mild winter like Boulder, Albuquerque, Las Vegas, Phoenix, and San Francisco, has a lower penalty than those cities for the same reason.

While the 15 cities for which we do our calculations represent a small fraction of global urban environments, they represent a wide range of climates, including 15 of the 16 ASHRAE climate zones specified for buildings in the US [60]. We therefore expect ARCs on similarly constructed buildings in similar climate zones, and experiencing similar wintertime weather, to incur similar penalties. For instance, based on current climates, ARCs on buildings in Riyadh may have a high penalty like observed in Phoenix, as the cities have comparable climate [61]. Likewise, Rio de Janeiro and Miami, or Seattle and London, may see similar penalties [61]. Although penalties in heat trapping are difficult to accurately predict for a specific city without modelling, based on Equations 2 and 3, it could be assumed that ARCs on roofs of similarly constructed buildings, in cities with similar wintertime ambient temperatures, total precipitable water levels and cloud covers, will likely have similar penalties. As far as generalizations of our findings are concerned, we limit ourselves to this conjecture, but note that in almost all of the scenarios we studied, ARCs have a significant negative climate impact relative to TRCs.

## Robustness of the Findings: Sensitivity Analysis

While the results in **Figure 3** were calculated for a specific set of scenarios, the general conclusion, that ARCs will have a net penalty relative to TRCs for the foreseeable future, appears robust. This is indicated by the penalties we see for 15 US cities representing diverse climate zones. We tested the robustness of our findings further, by performing sensitivity analyses of our findings to different simulation parameters. The specific details and results showing the time evolution of the benefits and penalties are presented in Section 5 of the Supporting Information. Here we present the net penalties. In brief, in all realistic scenarios, ARCs have a net penalty compared to TRCs for the foreseeable future.

One of the most important parameters for our sensitivity analysis was building insulation. We first note that our simulations were conservative in assuming modest insulation (R14) on roofs, which heightens the impact of TRC and ARC envelopes on the indoors, thus amplifying benefits. As shown in **Figure 5A**, net-penalties increase with increasing building insulation levels. This is because higher insulation (R38) reduces the benefits of ARCs by isolating the indoor environment from the roof envelope, while lower insulation (R6) does not increase it enough to overcome the penalties, except in very cold locations where low insulation would be impractical (Figure S19, Supporting Information).

We also saw similar trends for different ARC transition temperatures – for lower transition temperatures, the net penalty of ARCs over TRCs reduce, but only because ARCs act like TRCs for a greater fraction of the cold weather. Still, even for a considerably lower transition temperature of 15°C compared to the ~20°C for typical designs[17,18], the penalties are substantial, while a higher transition temperature (25°C) only heightens the net penalty due to prolonged operation in the heating mode (**Figure 5B**, and Figure S20 of the Supporting Information). As stated earlier, we also simulated ARCs that switch their solar reflectance, and both solar reflectance and thermal emittance. We found them to have net penalties for the foreseeable future as well.

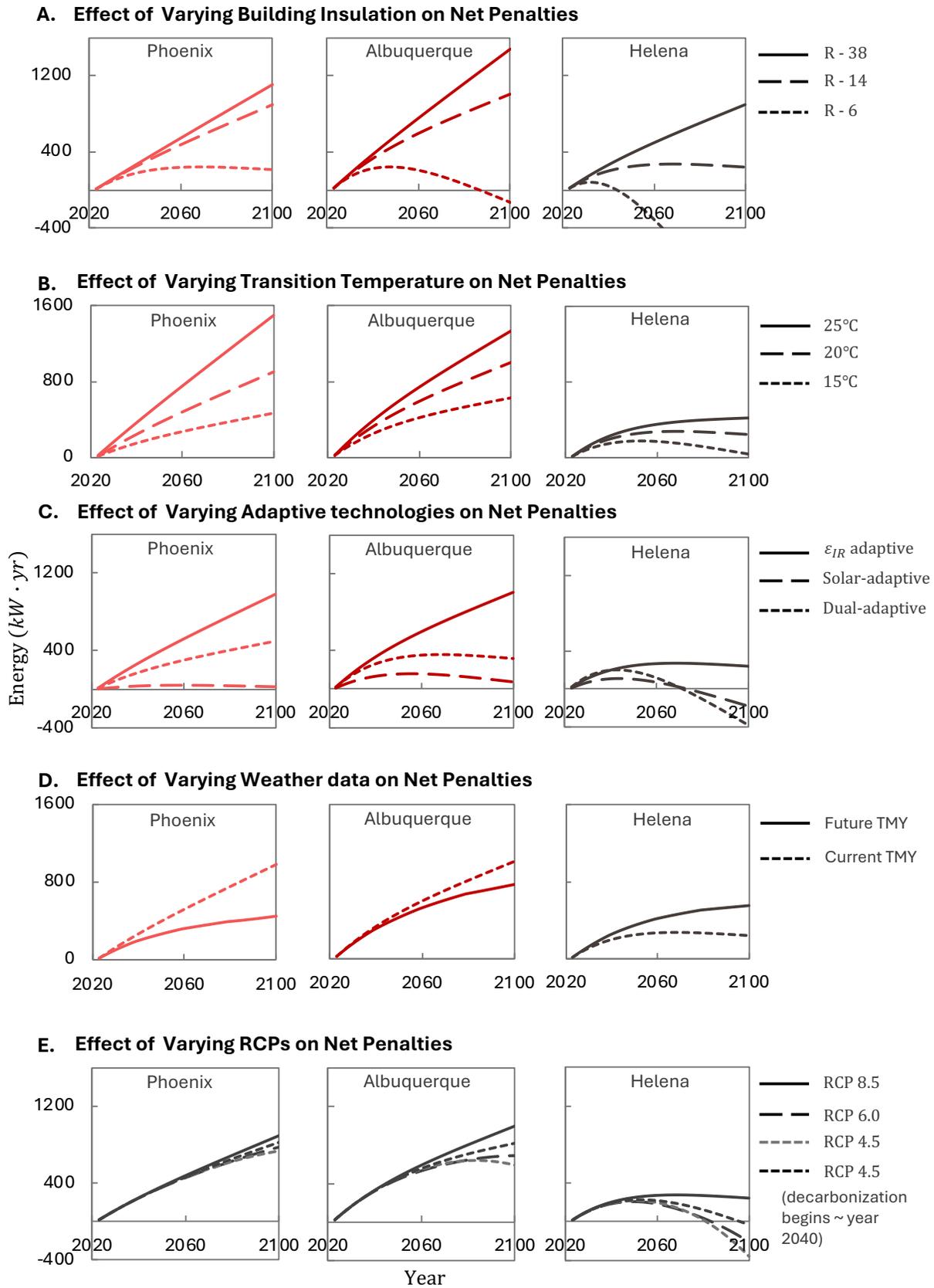

**Figure 5.** Time evolution of the net penalties accrued by an ARC over a TRC under variation of different parameters.

(**Figure 5C** and Figure S21 of the Supporting Information). One reason for the net penalties for solar-adaptive ARCs emitters being low is their effectiveness being limited to daytime hours. However, the strong heating effect of sunlight means that if paired with emittance switching, in colder regions like Helena, using a solar-adaptive or dual-adaptive design instead of a TRC would result in a net-benefit around 50 years post deployment.

We also explored different climate scenarios. Our primary focus was RCP 8.5, which assumes that $CO_2$ emissions follow current trends. For RCP 8.5, we considered two different climate projections, the one from Chowdhury et. al.[46] we used in **Figure 3**, and one where we assume that the climate continues to follow the current TMY up to 2100. Both yield net penalties for ARCs up to the end of the century (**Figure 5D**, and Figure S22 of Supporting Information). We also modelled net penalties for RCP 4.5 and 6.0 scenarios but note that those calculations are less accurate. This is because firstly, these scenarios assume global decarbonization that is not explicitly outlined for buildings, which meant that we could not model corresponding decarbonization of building energy usage. Secondly, the lack of TMY climate projections for RCP 4.5 and 6.0 forced us to rely on current TMY data. Nonetheless, we generally find net penalties persisting past this century, both when we assume building energy demands will be met by fossil fuels, and when we assume building energy decarbonization (**Figure 5E** and Figure S23 of Supporting Information). In fact, if building energy consumption is decarbonized – a likely scenario in the case of RCP 4.5 – it would reduce the benefit $\Delta Q_{greenhouse}$ of ARCs further, thereby increasing the net-penalties.

In this work, we focused on the impacts of ARCs and TRCs on individual buildings, whose small roofs do not appreciably impact the instantaneous temperature of the broader environment – the lower atmosphere essentially acts as a thermal reservoir and traps the heat. Thus, the penalty can be calculated for the roof alone. However, if deployed at scale, e.g. over all roofs of a city, radiative coolers could impact ambient temperatures,[62] thus changing the skywards radiosity of the broader environment as well. Although peripheral to our work, we considered the possibility where heat trapping by ARCs deployed at scale increases ambient temperatures relative to TRCs and thus, the heat the environment radiates to space (Supporting Information, Section 7). Our preliminary analysis shows that while the latter partially compensates the penalty of the ARCs, it is too small to have an appreciable impact (Supporting Information, Section 7). Thus, our findings are likely to hold for large scale deployment of ARCs over TRCs as well.

## Net Penalties, Conservative Estimates, and Operational Penalty as a Fundamental Limit

It is important to note that the aim of our study is not to calculate the precise penalties of deploying ARCs relative to TRCs on roofs. The precise amount would depend on factors like the immediate environment, sky-view factor, location-specific microclimate, and building design. Simulating such a wide variety of scenarios is beyond the scope of this work. Rather, our work is the first exploration of whether ARCs have a net thermal penalty relative to TRCs within reasonable accuracy, and whether that penalty is significant. The results in **Figure 3** and the sensitivity analysis in **Figure 4**, are affirmative on both fronts.

The values in **Figure 3** may, in fact, be conservative estimates. For instance, in calculating $\Delta Q_{rad,roof}$, we only considered the heat from TRCs and ARCs *directly transmitted* to space, which is lower than the actual heat lost.[57] For the same reason, in Equation 2, we do not consider the net heat lost to the clouds (which would be partially reemitted to the sky), which would heighten our penalty by a small amount. Likewise, our modelling of buildings with R14 insulation, even though buildings are increasingly better insulated, also heightens the benefits of ARCs. Section 6 of the Supporting Information details these aspects, and a few others, which suggest that ARCs may have net penalties that are considerably higher than what **Figure 3** shows.

We also note that our calculations do not account for dew, precipitation and dust on roofs, or long-term degradation of ARCs and TRCs. To our knowledge, reliable data on these aspects, which can be generalized to all TRCs or ARCs, do not exist. Nonetheless, we note that optical masking by water or dust, would likely reduce both relative penalties and benefits of ARCs.

A last, crucial point to note is that our study concerns the net *operational* thermal footprint of ARCs and TRCs. We do not consider the non-operational footprints of the cradle-to-grave journeys of these technologies, such as $CO_2$ emissions associated with manufacturing or (re)installation. This is deliberate. Firstly, our survey of the life cycles of representative ARCs and TRCs indicate that ARCs, being more sophisticated than TRCs, have multifold higher carbon emissions associated with their materials, manufacturing, and installation and maintenance (Supporting Information, Section 10). Our analysis, which are purely operational and do not take these into account, are therefore conservative in favor of ARCs.

The second reason is more fundamental. In focusing solely on the operational thermal penalty of ARCs relative to TRCs, and leaving out non-operational aspects, we show a fundamental limitation of ARCs, where the very functionality that yields the energy-efficiency and emissions reductions also leads to a net thermal penalty. This is a unique example of a negative relationship between energy-efficiency and climate impact. Moreover, it also represents a physical limit – ARCs will have an unavoidable thermal penalty relative to TRCs, regardless of the design of the ARCs or TRCs, and even if their non-operational carbon footprints dwindle with future technological advancements.

## Implications of Our Findings

### Energy Efficiency is not always related to a Positive Climate Impact

Discourse on sustainable built environments, whether it is regarding materials development, building design, or policymaking, are often dominated by energy-efficiency and emissions reductions.[63–68] Typically, energy-efficiency, which leads to emissions reductions, is understood to have a positive relationship with climate impact. By contrast, our work highlights the direct thermal impact of roofs through radiative heat loss to space and shows that in the context of ARCs vs TRCs, it has a greater impact than competing emissions reductions benefits. The results in **Figure 3** and **Figure 4** represent an important and critical example where energy-efficiency is negatively related with climate impact – something that has, to our knowledge, not been previously considered.

### Roof Envelopes – Material Design for TRCs, ARCs, and Other Static Emitters

Our study shows that ARCs on roofs, whether emittance-switching,[17,19,24] solar-adaptive,[18,19,21] or dual-mode,[20,69–72] would have thermal penalties or a negative climate impact for the foreseeable future. These results hold regardless of the switching mechanism of ARCs – whether it is electrochromic, thermochromic, mechanical, fluidic, or thermodynamic[17–22,24,72–76]. For building thermoregulation and design, this calls for a reconsideration of adaptive optical materials and strategies. We are mindful that decades of research has yielded outstanding ARCs that are either already being explored for roofs,[77–80] or increasingly seen as a way to sustainably thermoregulate buildings[23,35–40]. Thus, our findings are timely and important in light of current research.

Our findings are applicable beyond TRCs and ARCs to static optical envelopes as well. For instance, roofs with metal surfaces that operate on a low-$\varepsilon_{IR}$ mode all year have a higher $\Delta Q_{rad,roof}$ than $\varepsilon_{IR}$ switching ARCs, and they may also have a lower $\Delta Q_{greenhouse}$ because of their inability to lose heat to the sky and cool buildings in warm weather. Likewise, dark roofs, which trap solar heat on earth across seasons, and overheat buildings during warm weather, would have higher $\Delta Q_{rad,roof}$ and lower $\Delta Q_{greenhouse}$ than solar-adaptive ARCs. In other words, their climate impact relative to TRCs would be considerably worse than what we see for ARCs (Supporting Information, Table S4).

### Calculating the Net Operational Thermal Footprint of Building Envelopes

Beyond our study of radiative coolers, our consideration of both the direct radiative penalty and emissions reductions benefit (**Figure 2**) shows a more holistic way to calculate the operational thermal footprint of buildings using the optical properties of their envelopes. The framework we develop could be applied to any roof envelope, and with technical modifications we showed in a prior work [81], to vertical facades.[A] This is critical, because current metrics for evaluating green buildings focus on energy-efficiency and emissions reductions,[63–68] overlooking the direct radiative impact, which may cause envelope technologies like ARCs to be rated as environmentally friendly. Thus, our work shows a way to better quantify what makes a building 'green'. The concept can also be extended to radiant technologies, like photovoltaic panels, in general.

The above issues also may make our findings relevant to decision-making at government and organizational levels. In recent years, established TRCs like cool roof coatings, and new TRC [33,34,82–84] and adaptive coatings [85–87] have been promoted or accelerated towards adoption by governments [88,89] and the private sector [90,91]. As climate change mitigation and decarbonization goals become more relevant, our findings may inform policymaking in cases where trade-offs between energy-efficiency and climate impact are involved.

### Acknowledgements

J. M. conceived the research questions, developed the theoretical framework for the calculations, and supervised the research. N.V. performed the atmospheric radiative heat transfer modelling and quantifications of benefits and penalties of ARCs and TRCs. J. A. performed the building-level energy and thermal simulations, and WRF modelling. The authors acknowledge Dr Amilcare Porporato of Princeton University, and Dr Nadir Jeevanjee of NOAA Geophysical Fluid Dynamics Laboratory, for helpful discussions and advice. This study was supported by the startup funds from Princeton University SEAS.

### Declaration of Interest

J. M. is an inventor on patents WO/2019/113596 and PCT/US2016/038190, which detail ARCs and TRCs.

### Data Availability

All relevant data have been provided in the manuscript and Supporting Information. Additional request can be directed to the corresponding author.

---

[A] We speculate here that for vertical surfaces, specular and highly reflective surfaces across the solar to thermal infrared wavelengths may be better than TRCs or ARCs.